\documentclass[journal]{IEEEtran}
\pdfoutput=1
\usepackage{amsmath,amssymb, dsfont,bbm}
\usepackage{epsfig,latexsym,amssymb,amsmath,graphics,color}
\usepackage{graphicx,subfigure}
\usepackage{float}
\usepackage[linesnumbered,ruled]{algorithm2e}
\usepackage{epsfig,latexsym,amssymb,amsmath,graphics}
\usepackage{graphicx}
\usepackage[linesnumbered,ruled]{algorithm2e}
\usepackage{cite,color}
\usepackage{url,epstopdf}
\usepackage{stfloats}

\title{Wireless Communication Based on Microwave Photon-Level Detection With Superconducting Devices: Achievable Rate Prediction}
\author{Junyu Zhang, Chen Gong, Shangbin Li, Rui Ni, Chengjie Zuo, Jinkang Zhu, Ming Zhao, and Zhengyuan Xu
		\thanks{This work was supported by National Key Research and Development Program of China (Grant No. 2018YFB1801904), Key Program of National Natural Science Foundation of China (Grant No. 61631018), Key Research Program of Frontier Sciences of CAS (Grant No. QYZDY-SSW-JSC003). Junyu Zhang, Chen Gong, Shangbin Li, Chengjie Zuo, Jinkang Zhu, Ming Zhao, and Zhengyuan Xu are with Key Laboratory of Wireless-Optical  Communications, Chinese Academy of Sciences, School of Information Science and Technology, University of Science and Technology of China, Hefei, China. Email: jy970102@mail.ustc.edu.cn, \{cgong821,shbli, czuo, jkzhu, zhaoming, xuzy\}@ustc.edu.cn.

Rui Ni is with Huawei Technology, Shenzhen, China. Email: raney.nirui@huawei.com.}}
\date{}
\begin{document}
\maketitle{}

\begin{abstract}
Future wireless communication system embraces physical-layer signal detection with high sensitivity, especially in the microwave photon level. Currently, the receiver primarily adopts the signal detection based on semi-conductor devices for signal detection, while this paper introduces high-sensitivity photon-level microwave detection based on superconducting structure. We first overview existing works on the photon-level communication in the optical spectrum as well as the microwave photon-level sensing based on superconducting structure in both theoretical and experimental perspectives, including microwave detection circuit model based on Josephson junction, microwave photon counter based on Josephson junction, and two reconstruction approaches under background noise.
In addition, we characterize channel modeling based on two different microwave photon detection approaches, including the absorption barrier and the dual-path Handury Brown-Twiss (HBT) experiments, and predict the corresponding achievable rates. According to the performance prediction, it is seen that the microwave photon-level signal detection can increase the receiver sensitivity compared with the state-of-the-art standardized communication system with waveform signal reception, with gain over $10$dB.
\end{abstract}

\section{INTRODUCTION} \label{sec.Introduction}
With the development of future wireless communication system, the Internet of Things (IoT) technology that constitutes the “Internet of Everything” is promising for both industrial and daily applications. 
Thus, reducing the energy consumption of narrowband communication is attracting extensive interests from both academia and industrial areas. 
One key challenge is the signal detection under extremely weak power regime. 
Due to wave-particle duality of the electromagnetic field, the microwave waveform degrades into microwave photons under the extremely weak electromagnetic power. 
The signal detection need resort to the approach based on the microwave photon processing.

On the other hand, in optical spectrum, photon-level detection can be accomplished using the photoelectric effect, where the energy of a single photon is high enough to excite detectable electrons in room temperature. 
In microwave frequencies, the energy of a single microwave photon is lower to a magnitude of five orders, which increases the difficulties of photon-level detection, especially at room temperature with strong ambient radiation. 
Conventional detection adopts non-linear resistive components, typically Schottky barrier diodes as rectifiers and heterodyne mixers. 
The sensitivity of such mixer typically lies in the waveform power regime\cite{sis}.

In order to realize microwave photon-level detection in the microwave frequency band, it is necessary to reduce the background temperature such that the single photon energy is differentiable in the ambient radiation. 
The superconducting structure in low temperature serves as a good candidate for the receiver device, where electron-phonon interactions lead to weak gravitational forces between electrons. 
In some materials, at sufficiently low temperatures, such weak gravitational force can lead to a binding pair, i.e., a superconducting ground state. 
Single-element excitation above this ground state requires a minimum threshold energy and consists of quasi-particles with both “electrons” and “holes”. 
Cohen, Falicov, and Phillips proposed the Hamiltonian theory to characterize the potential barrier between quasi-particle tunneling between two volume superconductors\cite{hamidun}. 
After that, Josephson made an significant prediction that the superconducting pair could also pass through the barrier tunnel and derived the consequences of the “Josephson Tunnel” using the same Hamiltonian model. 
The Josephson junction consists of two weakly connected superconductors, such that a pair of currents can only be generated on both sides of the Josephson junction by tunneling through a weak barrier. \cite{sis}
A typical model of the Josephson junction is a superconducting-insulator-superconductor (SIS) interlayer.

When the quantum voltage at the incident frequency exceeds the voltage width of the quasi-particle tunneling threshold, the SIS responds to a single quantum through a photon-assisted tunneling mechanism. This allows the SIS junction to perform microwave/millimeter wave photon detection close to the quantum limit\cite{SIS1,SIS2,SIS3,SIS4}. Besides, a single microwave photon can also be detected by microwave-mechanical-optical coupling\cite{guangji1,guangji2,guangji3,guangji4,guangji5,guangji6} or nitrogen-vacancy center\cite{NV1,NV2}, but the superconducting method still has the advantages of high sensitivity and standardized manufacturing process.

The strong interaction of superconducting integrated circuits with microwave photons forms the basis of circuit quantum electrodynamics (cQED). An efficient and versatile microwave photon counter can be realized based on microwave-induced transitions between discrete energy levels in the Josephson junction \cite{counter}. Utilizing more efficient counter and microwave photon detection element \cite{dianlv}, we can perform various types of experiments and measurements, including Bell's inequality experiments, all quantum based on optics and measurement, quantum computation, quantum homodyne tomography, and quantum communication and cryptography\cite{exp1,exp2}. These are essential for quantum information processing and communication \cite{hbt}.

Particularly in superconducting circuits, photoelectric conversion can be accomplished by the coupling of optical and microwave fields using micromechanical resonators \cite{airtoforce}, where the microwave radiation collected from the target area is phase conjugated and upconverted into a light field. Such kind of microwave quantum illumination system or quantum radar outperforms conventional microwave radar\cite{airtoelc}.

In this paper, we firstly propose the concept of quantumized signal reception architecture based on superconducting devices. Then, we overview existing theoretical and experimental works on photon-level microwave signal reception. We predict the achievable transmission rates for two types of superconducting structure, based on absorption barrier and two-path Handury Brown-Twiss (HBT) experiments. It is predicted that compared with the state-of-the-art communication standard, the proposed approach can outperform via over $10$dB.  

The remainder of this work is organized as follows. In Section \ref{sec.sys}, we outline the system architecture with superconducting signal reception device. In Section \ref{sec.Review of Existing Works}, we overview the existing works on the superconducting structure based signal reception. In Section \ref{sec.SAB}, we analyze the achievable rate based on absorption barrier. In Section \ref{sec.HBT},  we analyze the achievable rate based on two-path signal measurement. Finally, concluding remarks are given in Section \ref{sec.CONCLUSION}.

\section{SYSTEM ARCHITECTURE}\label{sec.sys}
\begin{figure*}[htbp]
  \setlength{\abovecaptionskip}{-0.2cm} 
  \setlength{\belowcaptionskip}{-2cm}
  \centering
  \includegraphics[width=2\columnwidth]{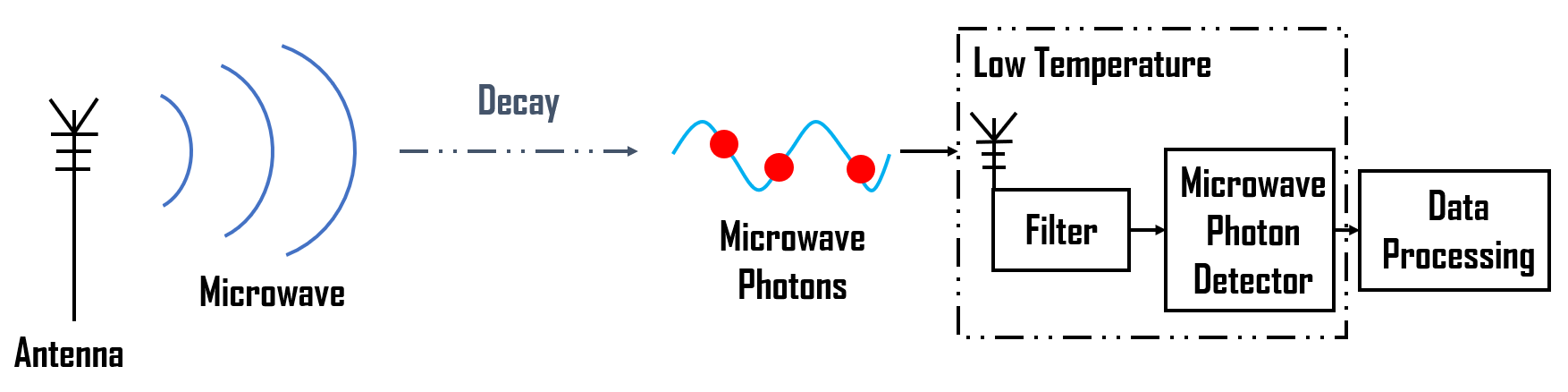}
  \caption{System architecture of wireless communication transmission based on quantumized microwave photon-level detection.}
  \label{sys_1}
\end{figure*}

In this section, we propose a system architecture for wireless communication based on quantumized microwave photon-level detection with superconducting
devices, as shown in Figure \ref{sys_1}. 

The transmitting antenna sends the modulated signal, which suffers strong attenuation during the transmission in the space. According to the wave-particle duality, the microwave signal exhibits a particle image, which calls for quantumized microwave photon-level signal detection.
At the receiver side, we adopt a microwave photon-level detector based on superconducting devices for signal reception, and perform data processing based on its output signal. A narrowband filter is adopted ahead of the photon-level detector to reduce the interference in the neighborhood spectrum.

\section{REVIEW OF EXISTING WORKS} \label{sec.Review of Existing Works}

We overview existing works and basic principles on photon-level detection in the optical spectrum, as well as the microwave photon-level signal detection, including microwave photon counters based on superconducting structures and the reconstruction of quantum states in a noisy environment.

\subsection{Photon-level Communication System in the Optical Spectrum}
The photon-level signal detection-based communication system design has been extensively studied for optical communication.  Within a fixed time period, the number of detected photons conforms to the following Poisson distribution,
\begin{equation}
P(\mathrm{N}=n)=\frac{\bar{\lambda}^{n}}{n !} e^{-\bar{\lambda}},
\end{equation}where the mean number $\bar{\lambda}$ can be obtained from the received signal energy within that period divided by the energy per photon. The achievable communication rate and signal detection can be performed based on the Poisson distibuted number of photons.

For such types of channels, it can be proved that on-off keying (OOK) modulation with sufficiently short symbol interval can approach the capacity \cite{info1sec3,info2sec3,info3sec3}. More specifically, letting $\lambda_s$ and $\lambda_b$ denote the mean photon numbers of the signal and noise components, respectively, the mean numbers of photons corresponding to symbols one and zero are $\lambda_s + \lambda_b$ and $\lambda_b$, respectively. The signal detection performance can be analyzed based on the following two Poisson distributions for symbol zero and symbol one,
\begin{equation}P(n | \text {zero})=\frac{\lambda_{b}^{n}}{n !} e^{-\lambda_{b}},\end{equation}
\begin{equation}P(n | \text {one})=\frac{\left(\lambda_{\mathrm{s}}+\lambda_{b}\right)^{n}}{n !} e^{-\left(\lambda_{\mathrm{s}}+\lambda_{b}\right)}.\end{equation}

In the atmosphere, the above photon-level signal model can well characterize the non-line-of-sight(NLOS) ultraviolet scattering communication, where large path loss of the NLOS link leads to photon-level signal intensity and negligible background radiation can guarantee the photon-level signal sensitivity. The inter-symbol interference structure due to the finite bandwidth transmitter device is investigated in \cite{ISI}. To expand the transmisson range, the multi-hop relay communication has been investigated in \cite{ISI4,XX3sec3a}.

Under a turbulence channel, the received signal exhibits a doubly stochastic characteristic due to both signal intensity variation and Poisson distributed photons given fixed signal intensity. Such type of channel can be characterized by a hidden Markov chain model, where the hidden state represents the strength of the received signal, and the observation represents the number of photons given the hidden state. Assume that there are $K$ states in the mixed Poisson signal model. The received signal can be characterized by the following mixed Poisson model,
\begin{equation}P(N=n)=\sum_{k=1}^{K} p_{k} \frac{\lambda_{k}^{n}}{n !} e^{-\lambda_{k}},\end{equation}where $p_k$ and $\lambda_k$ denote the probability and mean number of photons corresponding to state $k$, respectively. The related parameters of the probability distribution can be estimated via expectation-maximization(EM) algorithm \cite{gujic1}.

Due to the finite bandwidth of the receiver detector, the output pulse of each photon exhibits positive rising edge and falling edge. Such phenomenon causes positive dead time,  leading to the mergence of the two pulses if the incidence time between the two photons is shorter than the dead time. The received signal statistics and symbol detection are characterized in \cite{jiesh1}.

For the photon-level communication system realization, the receiver performs counting-based signal synchronization, Poisson distribution-based signal detection and channel code decoding. More details on the system realization and experiments can be found in \cite{jiesh3}.

\subsection{Microwave Photon Counter Based on Josephson Junction}
\begin{figure}[htbp]
  \setlength{\abovecaptionskip}{-0.2cm} 
  \setlength{\belowcaptionskip}{-2cm}
  \centering
  \includegraphics[width=0.8\columnwidth]{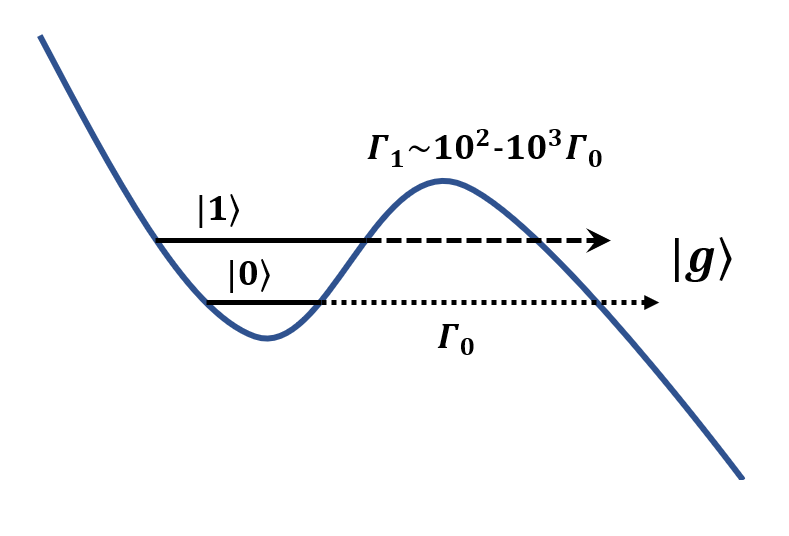}
  \caption{Junction potential energy landscape. }
  \label{sec31}
\end{figure}

According to \cite{hbt}, as the bias current of Josephson junction approaches a critical value $I_0$, the potential of the junction approaches the local minimum, which consists of a few discrete energy levels. The resonance of receiving a microwave photon produces an oscillation between ground state $| 0 \rangle$ and excited state $| 1 \rangle$ of the Josephson junction, which tunnels to a continuous state at rates $\Gamma_0$ and $\Gamma_1$, respectively. Since the tunneling rate depends on the barrier height, $\Gamma_{1}$ tends to be 2-3 orders of magnitude larger than $\Gamma_{0}$, as shown in Figure \ref{sec31}. Due to different tunneling rates for the two states, a microwave photon-counter can be designed.

\begin{figure}[htbp]
  \setlength{\abovecaptionskip}{-0.2cm} 
  \setlength{\belowcaptionskip}{-2cm}
  \centering
  \includegraphics[width=0.8\columnwidth]{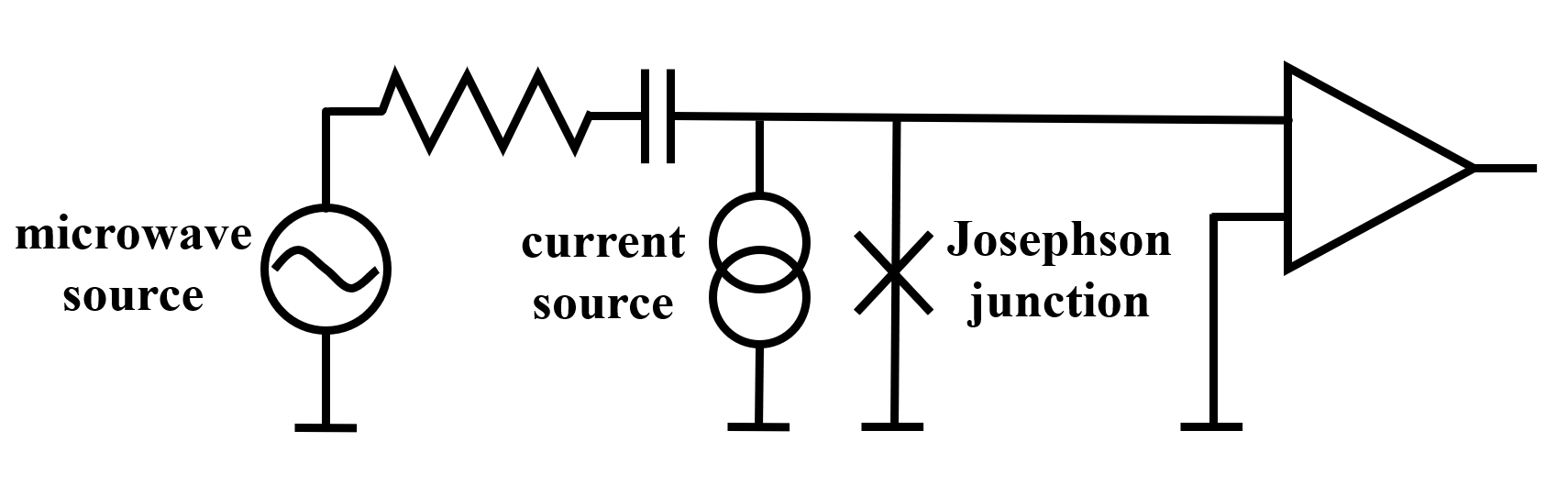}
  \caption{Schematic diagram of the circuit.}\label{sec31dian}
\end{figure}

Based on this principle, the direct-current (DC) operating point of the microwave counter is set to be close to the critical current of Josephson junction. The peripheral circuit of the Josephson junction can almost resonate with the incident microwave photon, causing the transition from ground state $| 0 \rangle$ to excite state $| 1 \rangle$. The Josephson junction in the excite state has a faster tunneling rate than the ground state, leading to a large pulse, and the comparator tends to output a count. The schematic diagram of the circuit is shown in Figure \ref{sec31dian}.

To check the performance of the counter, we define its efficiency
\begin{equation}
\eta \equiv P_{\text {bright}}(1-P_{\text {dark}}),
\end{equation}
where $P_{\text { bright }}$ denotes the probability that the comparator can correctly generate the count after the incidence of the microwave counter; and $P_{\text {dark}}$ denotes the probability that the Josephson junction generates a count under no microwave incidence.

It is reported that such efficiency can reach 0.7 and can be kept above 0.6 in a 450 MHz wide frequency band\cite{hbt}.
\subsection{A Microwave Photon Counter Based on a Discrete Superconducting Structure}
A microwave photon counter based on discrete superconducting structure employs metamaterial that realizes single-wave photon detection through irreversible absorption of microwave photons\cite{xsq}. When a microwave photon enters the device, it is captured with a certain probability. Then, the system reachs a stable state that can be observed mesoscopically after the absorption process. More specifically, the microwave photon counter consists of a set of photon absorbers placed along an one-dimensional waveguide, which can be constructed with a bi-stable quantum circuit similar to that used to implement the qubit. These circuits are capable of capturing photons and transitioning from initial state $|0\rangle$ to steady state $|g\rangle$.

In the counting process, the counting is done after the absorption process. The detection process is thus passive and does not require any extra control. Numerical results show that as the number of photon absorbers increases, the entire photon absorption/detection probability can reach $90\%$\cite{xsq}.

\subsection{Microwave Quantum State Reconstruction in Noisy Environment}
For microwave communication based on photon-level detection, information can be detected through the quantum states at the receiver, which inevitably suffers noise and needs to be reconstructed based on statistical methods.
\subsubsection{Conjugate Orthogonal Component Quantum State Reconstruction}

Consider the quantum states of field mode $a$, which can be reconstructed using infinite set of moments $\left\langle\left(a^{\dagger}\right)^{n} a^{m}\right\rangle$ for $m, n \geq 0$. The mark $'\dagger'$ stands for Hermitian conjugate.  Moreover, it can be approximated by finite terms via limiting the size of $n+m$\cite{oneroad}.

Assume that $a$ is amplified by a linear amplifier with factor $G$, which introduces noise $h$. The amplified signal is mixed with in-phase and out-of-phase local oscillators in the mixer to perform frequency-down processing to detect a set of conjugate orthogonal components $\widehat{X}$ and $\widehat{P}$, given by
\begin{equation}
\sqrt{G}(\hat{X}+i \hat{P})=\sqrt{G}\left(a+h^{\dagger}\right) \equiv \hat{S}.
\end{equation}

\begin{figure}[htbp]
  \setlength{\abovecaptionskip}{-0.2cm} 
  \setlength{\belowcaptionskip}{-2cm}
  \centering
  \includegraphics[width=0.9\columnwidth]{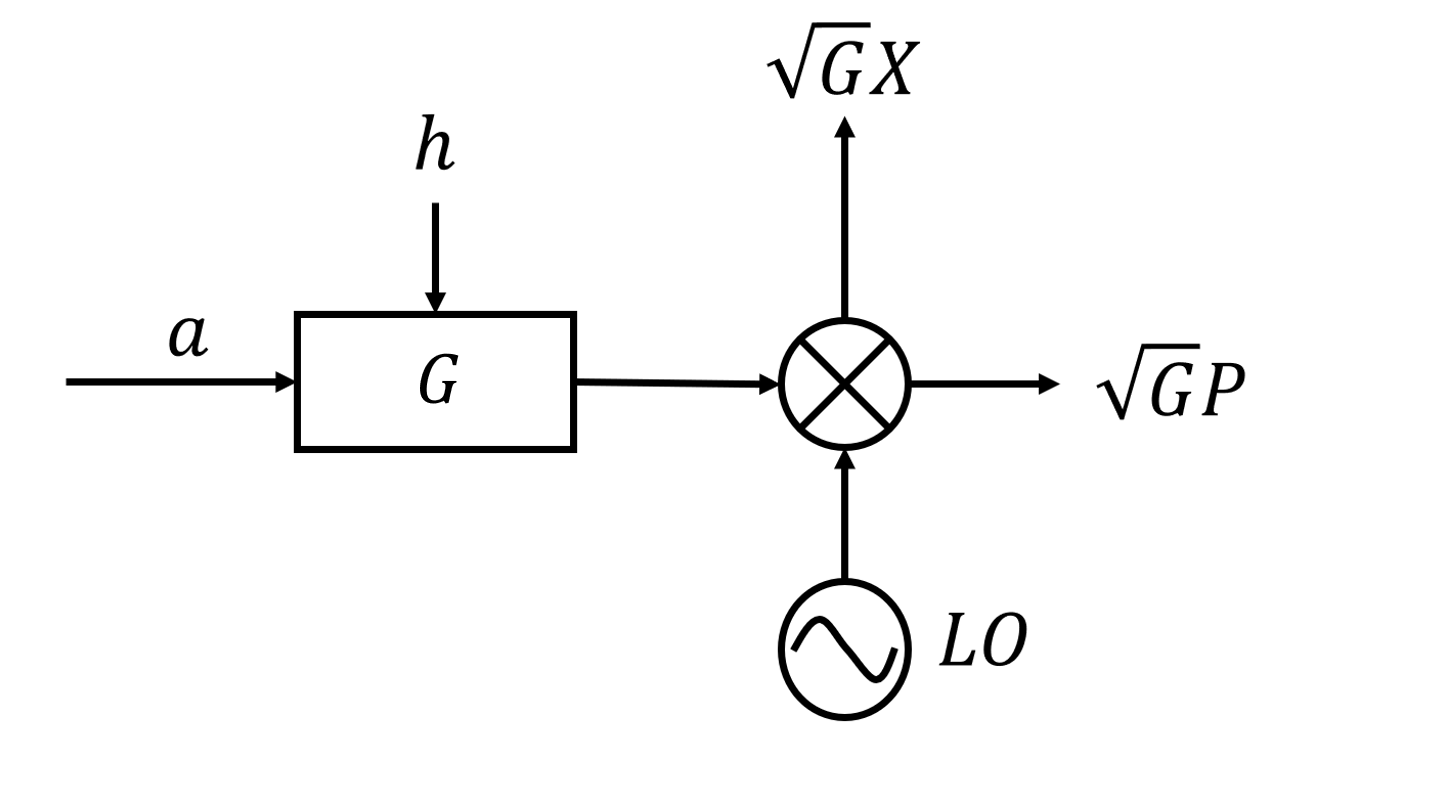}
  \caption{Generation of orthogonal components $\hat{X}$ and $\hat{P}$. }
  \label{secfig331}
\end{figure}

Noise $h$ can be well approximated as a Gaussian phase space distribution. Besides, the measured distribution of $\hat{S}$ at the output is
\begin{equation}
D^{[\rho]}(S)=\frac{1}{G} \int d^{2} \beta P_{a}(\beta) Q_{h}\left(\frac{S^{*}}{\sqrt{G}}-\beta^{*}\right),
\end{equation}
where $P_a$ is the Glauber-Sudarshan P function of mode $a$ and $Q_h$ is the Husimi Q function of noise $h$ \cite{Pf1,Pf2,Qfunction}. Figure \ref{secfig331} shows the process of obtaining the amplified expectations of two orthogonal components $\hat{X}$ and $\hat{P}$ from the input signal. In particular, for vacuum state $| 0 \rangle$, $P_{a}(\beta)=\delta^{(2)}(\beta)$, and thus
\begin{equation}
D^{[ | 0\rangle\langle 0 |]}(S)=\frac{1}{G} Q_{h}\left(\frac{S^{*}}{\sqrt{G}}\right).
\end{equation}

The second state $| \psi \rangle$ in the signal detection can be selected as the state of interest, such as Fock state $| 1 \rangle$. Then, we can get $D^{[ | \psi\rangle\langle\psi |]}$ by measurement, and thus
\begin{equation}
\langle(\hat{S}^{\dagger})^{n} \hat{S}^{m}\rangle_{\rho}=\int d^{2} S\left(S^{*}\right)^{n} S^{m} D^{[\rho]}(S).
\end{equation}

When the signal and noise are not correlated, we have
\begin{equation}
\begin{aligned}	 
	 \langle(\hat{S}^{\dagger})^{n} \hat{S}^{m}\rangle_{\rho} =  & G^{(n+m) / 2} \sum_{i, j=0}^{n, m} C_m^j C_n^i  \left\langle(a^{\dagger})^{i} a^{j}\right\rangle \\
	 & \times \langle h^{n-i} (h^{\dagger})^{m-j}\rangle.
\end{aligned}
\end{equation}
More specifically, for  a vacuum state, we have
\begin{equation}
\langle(\hat{S}^{\dagger})^{n} \hat{S}^{m}\rangle_{| 0 \rangle\langle 0 |}=G^{(n+m) / 2}\langle h^{n-i} (h^{\dagger})^{m-j}\rangle.
\end{equation}

Based on $\langle(\hat{S}^{\dagger})^{n} \hat{S}^{m}\rangle_{\rho}$ and $\langle(\hat{S}^{\dagger})^{n} \hat{S}^{m}\rangle_{ | 0 \rangle\langle 0 |}$,
we can get $\left\langle\left(a^{\dagger}\right)^{n} a^{m}\right\rangle$, such that the quantum state reconstruction can be completed. Based on $\left\langle\left(a^{\dagger}\right)^{n} a^{m}\right\rangle$, we can restore the information on the quantum state, e.g., the following Winger function of state $S$,

\begin{equation}
\begin{aligned}	 
W(\alpha)  = &\sum_{n, m} \int d^{2} \lambda \frac{\langle(a^{\dagger})^{n} a^{m}\rangle(-\lambda^{*})^{m} \lambda^{n}}{\pi^{2} n ! m !}  e^{-\frac{1}{2}|\lambda|^{2}+\alpha \lambda^{*}-\alpha \lambda^{*} \lambda}.
\end{aligned}
\end{equation}

\subsubsection{Dual Signal Reconstruction Scheme}

A two-way reconstruction scheme that obtains the statistics $\left\langle S^{n}\right\rangle$ of signal S is reported in \cite{tworoad}. Assuming signal $\mathrm{S} \sqrt{2}$ is equally separated by a four-port 50-50 microwave beam splitter, which are amplified via $G$ with noise components $\chi_{1}$ and $\chi_{2}$. The output signals, denoted as $C_1$ and $C_2$, are given by
\begin{equation}
\begin{array}{c}{C_{1}=\mathrm{G}\left(\mathrm{S}+\mathrm{V}+\chi_{1}\right)}, \\ {C_{2}=\mathrm{G}\left(-\mathrm{S}+\mathrm{V}+\chi_{2}\right)},\end{array}
\end{equation}
\begin{figure}[htbp]
  \setlength{\abovecaptionskip}{-0.2cm} 
  \setlength{\belowcaptionskip}{-2cm}
  \centering
  \includegraphics[width=0.9\columnwidth]{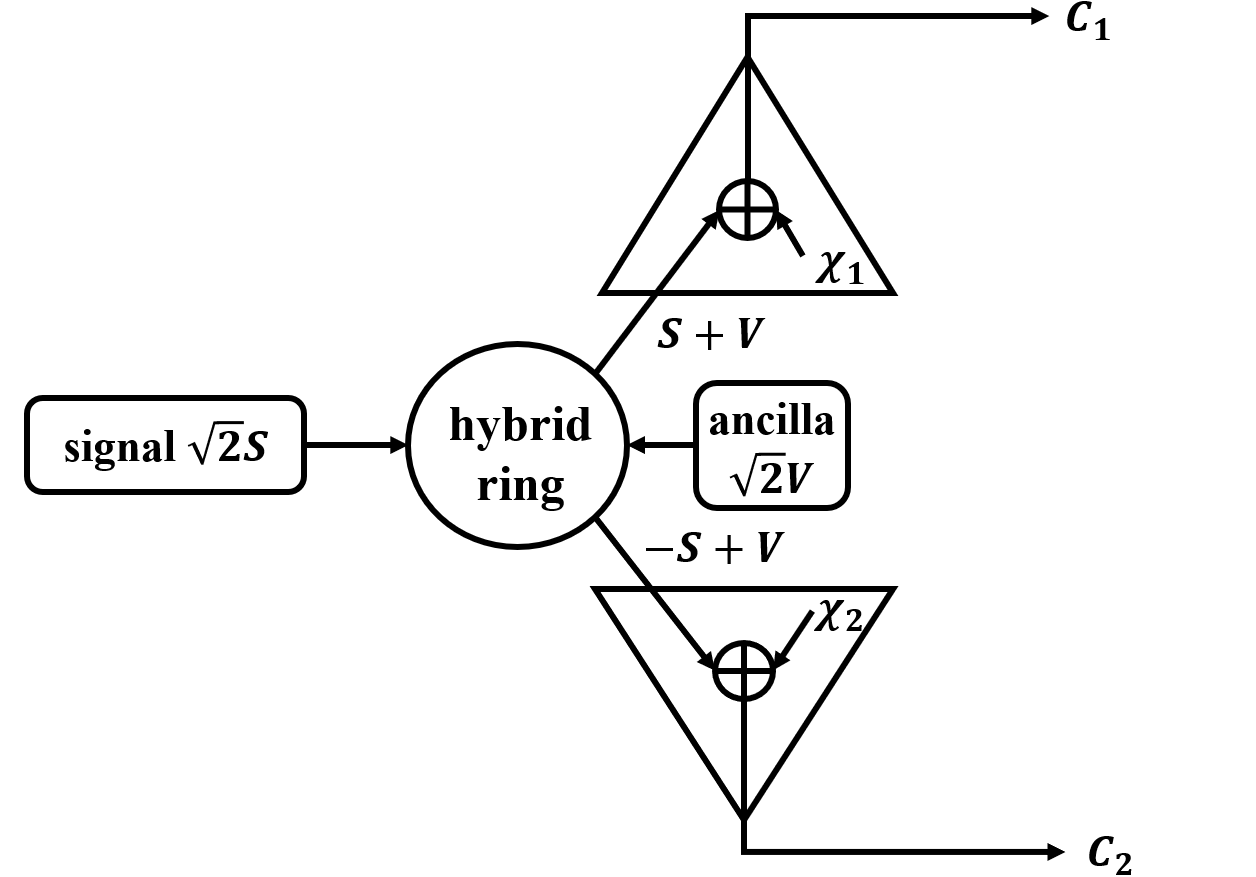}
  \caption{Dual signal model.}\label{dualfig} 
  
\end{figure}
where $V$ is an additional signal artificially introduced from the four-port beam splitter, assuming that the information of $V$ is completely known. Figure \ref{dualfig} shows the process from the input signal to dual signal. It is assumed that $\left\langle\chi_{1}\right\rangle=\left\langle\chi_{2}\right\rangle= 0$, $\left\langle V^{2 j+1}\right\rangle= 0$, and that the value of $\left\langle C_{1}^{l} C_{2}^{m}\right\rangle$ can be obtained by measurement, such that the statistics of signal $S$ and noise $\chi$ can be obtained. Then, we have that
\begin{equation}
\begin{aligned}\langle S^{n}\rangle&=\langle C_{1}^{n-1} C_{2}\rangle / G^{n} \\ &-\sum_{k=1}^{n-1} \sum_{j=0}^{k} C_{n-1}^{k}C_{k}^{j}\langle S^{n-k}\rangle\langle V^{j}\rangle\langle\chi_{1}^{k-j}\rangle \\ &+\sum_{k=0}^{n-1} \sum_{j=0}^{k}C_{n-1}^{k}C_{k}^{j}\langle S^{n-k-1}\rangle\langle V^{j+1}\rangle\langle\chi_{1}^{k-j}\rangle \end{aligned}
\end{equation}

\begin{equation}
\begin{aligned}
\langle\chi_{1}^{n}\rangle=&+ \langle C_{1}^{n}\rangle / G^{n}\\
&-\sum_{k=1}^{n} \sum_{j=0}^{k} C_n^k C_k^j \langle\chi_{1}^{n-k}\rangle\langle S^{k-j}\rangle\langle V^{j}\rangle,
\end{aligned}
\end{equation}

\begin{equation}
\begin{aligned}
\langle\chi_{2}^{n}\rangle =& +\langle C_{2}^{n}\rangle / G^{n}\\
&-\sum_{k=1}^{n} \sum_{j=0}^{k} C_n^k C_k^j (-1)^{k-j}\langle\chi_{2}^{n-k}\rangle\langle S^{k-j}\rangle\langle V^{j}\rangle.
\end{aligned}
\end{equation}

It can be seen that the signal statistics can be obtained by a recursive method. The central moment can be obtained from:
\begin{equation}
\left\langle(S-\langle S\rangle)^{n}\right\rangle=\sum_{k=0}^{n} C_n^k (-1)^{n-k}\left\langle S^{k}\right\rangle\langle S\rangle^{n-k}.
\end{equation}

\section{MIRCOWAVE PHOTON DETECTION BASED ON SUPERCONDUCTING ABSORBING BARRIER}\label{sec.SAB}

We elaborate the model of superconducting absorption barrier\cite{xsq}, outline the communication system model, and numerically evaluate the achievable rate.

\subsection{Superconducting Absorption Barrier \cite{xsq}}
The superconducting absorption barrier consists of a set of microwave photon absorbers coupled by a one-dimensional waveguide\cite{xsq}, which can be realized by a Josephson junction at a critical current bias and is able to capture and count the microwave photons.

When the Josephson junction is close to the critical current bias, the local minimum of potential consists of discrete energy levels, such as the ground state $|0\rangle$ and excite state $|1\rangle$. States $|0\rangle$ and $|1\rangle$ transition to the continuous state $|g\rangle$ at the rates of $\Gamma_{0}$ and $\Gamma$, respectively, where $\Gamma \gg \Gamma_{0}$.

The Hamiltonian $H$ contains the absorber and radiation field, as well as wave functions $\psi_{l}$ and $\psi_{r}$ propagating in the left and right directions at group velocity $v_g$, respectively. The interaction between the two wave functions is established by the delta potential field of intensity $V$, given by
\begin{equation}
\begin{aligned} 
H =& \sum_{i} \hbar \omega | 1 \rangle_{i}\langle 1 | +i \hbar v_{g} \int d x S_{lr} \\ 
&+\sum_{i} \int d x V \delta(x-x_{i})[\psi_{lr} | 1\rangle_{i}\langle 0 |+H . c .], \end{aligned}
\end{equation}
where $S_{lr}=\psi_{l}^{\dagger} \partial_{x} \psi_{l}-\psi_{r}^{\dagger} \partial_{x} \psi_{r}$, $\psi_{lr}=\psi_{l}+\psi_{r}$, $x_i$ and $| 0 \rangle_i$, $| 1 \rangle_i$ are the position and states of the $i^{th}$ absorber, respectively.

The density matrix of the waveguide and absorber is characterized by
\begin{equation}
\frac{\partial \rho}{\partial t}= -\frac{i}{\hbar}[H, \rho]+\mathcal{L} \rho,
\end{equation}
where $\mathcal{L}=\sum_{i} \mathcal{L}_{i}$; and ${\cal L}_i$ represents the standard attenuation term for each absorbed microwave photon $i$, given by
\begin{equation}
\mathcal{L}_{i} \rho=\frac{\Gamma}{2}(2|g\rangle_i\langle 1|\rho| 1\rangle_{i}\langle g|-| 1\rangle_{i}\langle 1|_{i} \rho-\rho | 1\rangle_{i}\langle 1|).
\end{equation}

For each absorber $j$, the quantum state transition can be characterized by the incident intensity, reflection $|r|^2$, and transmission intensity $|t|^2$ , given by the following scattering matrix ${\bold T}_j$,
\begin{equation}
{\bold T}_j=\left( \begin{array}{cc}{1-1 / \gamma} & {-1 / \gamma} \\ {1 / \gamma} & {1+1 / \gamma}\end{array}\right),
\end{equation}
where $\gamma=(\Gamma-\sqrt{-1} \delta) v_{g} / V^{2}$, $\delta=\omega-\omega_{\mu}$, and $\omega_{\mu}$ is the characteristic frequency of the absorber. For each absorber, absorption is equivalent to microwave photons that are not reflected or transmitted, such that the absorption rate of a single absorber is given by
\begin{equation}
\alpha=1-\frac{1+\left|T_{01}\right|^{2}}{\left|T_{11}\right|^{2}}=\frac{2 \gamma}{(1+\gamma)^{2}}.
\end{equation}

\begin{figure}[htb]
  \centering
  \includegraphics[width=0.5\textwidth]{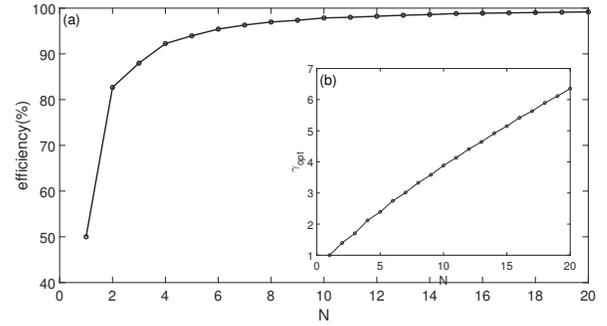}
  \caption{(a)The relationship between the optimal absorption efficiency and the number N of absorbers, (b)The relationship between the optimal parameter values and the number N of absorbers.}
  \label{xsq_p}
\end{figure}

It can be seen that the absorption probability of a single absorber is at most 0.5, while the absorption rate of the whole system can be increased by adopting multiple absorbers. Via increasing the number of absorbers $N$ and optimizing parameter $\gamma$, the absorption probability above 90$\%$ can be achieved, as shown in Figure \ref{xsq_p}.

For real circuit, assume that the capacitance of Josephson junction is $C_j$, the gate capacitance of the Josephson junction frontend and the microwave coupling is $C_g$. Then, parameter $\gamma$ can be expressed as
\begin{equation}
\gamma=\frac{\alpha^{2}}{c_{12}} \frac{\hbar}{e Z_{0}} \frac{\Gamma_{1}-i\left(\omega-\omega_{\mu}\right)}{\omega_{\mu}},
\end{equation}
where $c_{12}=C_{g} /\left(C_{g}+C_{j}\right); \alpha^{2}=4 e^{2} / C_{j} \hbar \omega$; $\Gamma_{1}$ is the rate at which $| 1 \rangle$ is converted to $| g \rangle$; and $Z_0$ is a constant.

\subsection{Equivalent Poisson Channel Modeling}
Refer to the system model shown in Figure \ref{sys_1}. Consider time-varying transmission power from on-off keying (OOK) modulationat the transmitter side with peak power constraint. Due to strong channel attenuation in the communication link, the received signal exhibits discrete number of signal photons yielding time-varying Poisson distribution, where the mean number of photons per slot is $\lambda_1$ for symbol on, and the intensity of background photon is $\lambda_{0}$.

\begin{figure}[htbp]
  \setlength{\abovecaptionskip}{-0.2cm} 
  \setlength{\belowcaptionskip}{-2cm}
  \centering
  \includegraphics[width=\columnwidth]{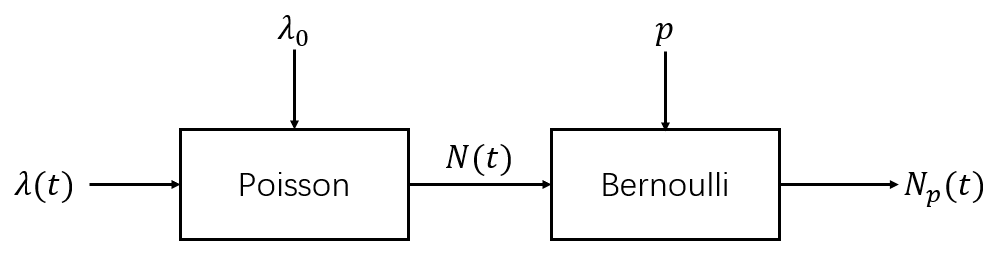}
  \caption{ Poisson channel with probability of reception. }
  \label{Poisson_x2}
\end{figure}

For channels based on superconducting absorption barrier, assume that each photon is absorbed by the barriers independently with probability $p$. Thus, the entire system model is given by Figure~$\ref{Poisson_x2}$.

\begin{figure}[htbp]
  \setlength{\abovecaptionskip}{-0.2cm} 
  \setlength{\belowcaptionskip}{-2cm}
  \centering
  \includegraphics[width=\columnwidth]{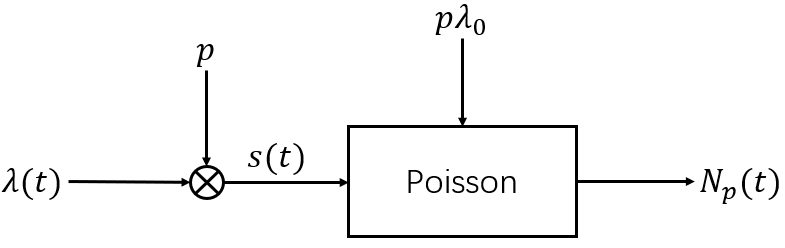}
  \caption{The equivalent Poisson channel model.}
  \label{Poisson_x3}
\end{figure}

Thus, we have $N(t) \sim \mathcal{P}((\lambda_1+\lambda_{0})\tau)$ and $N_{p}(t) \sim \mathcal{P}(p (\lambda_1+\lambda_{0})\tau)$, where $P(\lambda)$ denotes Poisson distribution with mean $\lambda$. Due to random extraction property of Poisson distribution, the transmission system can be characterized in Figure \ref{Poisson_x3}.

For simplicity, we consider hard-decision based on the output signal. Let $M$ denote the detection threshold, such that symbol on is detected of the number of detected photons exceeds $M$. Let 
\begin{equation}
P_{1}=\sum_{k=M}^{\infty} e^{-p\left(\lambda_1+\lambda_0\right)\tau} \frac{p^{k}\left(\lambda_1+\lambda_0\right)^{k}}{k !},
\end{equation}
\begin{equation}
P_{2}=\sum_{k=0}^{M-1} e^{-p \lambda_{2}\tau} \frac{p^{k} \lambda_{2}^{k}}{k !}.
\end{equation}

Assume that symbol one is transmitted with probability $\gamma$, i.e., $P(X = 1) = \gamma$ and $P(X = 0) = 1 - \gamma$. We have the following conditional probability,
\begin{equation}
\begin{aligned} \mathrm{P}(\mathrm{Y}=1 | \mathrm{X}&=1 )=P_{1}, \mathrm{P}(\mathrm{Y}=0 | \mathrm{X}=1)=1-P_{1}, \\ \mathrm{P}(\mathrm{Y}=1 | \mathrm{X}&=0 )=P_{2}, \mathrm{P}(\mathrm{Y}=0 | \mathrm{X}=0)=1-P_{2}, \end{aligned}
\end{equation}
where $P(\mathrm{Y}=1)=\gamma P_{1}+(1-\gamma)\left(1-P_{2}\right)$. For simplicity, we consider the following mutual information for the binary asymmetric channel,
\begin{equation}
\mathrm{I}(\mathrm{X}, \mathrm{Y})=\mathrm{H}\left(P_{\gamma}\right)-\gamma \mathrm{H}\left(P_{1}\right)-(1-\gamma) \mathrm{H}\left(P_{2}\right),
\end{equation}
where function $H(x) = -xlog_2x - (1-x)log_2(1-x)$. For the binary asymmetric channel under consideration, the optimal distribution on output $Y$ is given by 
\begin{equation}
P_{\gamma}^*=\frac{1}{1+2^{\frac{H\left(P_{1}\right)-H\left(P_{2}\right)}{P_{1}+P_{2}-1}}}.
\end{equation}

Moreover, higher achievable rate can be obtained via directly computing from the two Poisson distributions without hard decision between them.
\subsection{Simulation Result}\label{sec.SAB_sum}
Assume that the single photon absorption probability is 0.9, and that the code rate is 1 kbps (symbol period 1ms). The carrier frequency of the transmitted microwave photon is $5 \mathrm{GHz}$.  

\begin{figure*}[htbp]
  \setlength{\abovecaptionskip}{-0.2cm} 
  \setlength{\belowcaptionskip}{-2cm}
  \centering
  \includegraphics[width=2\columnwidth]{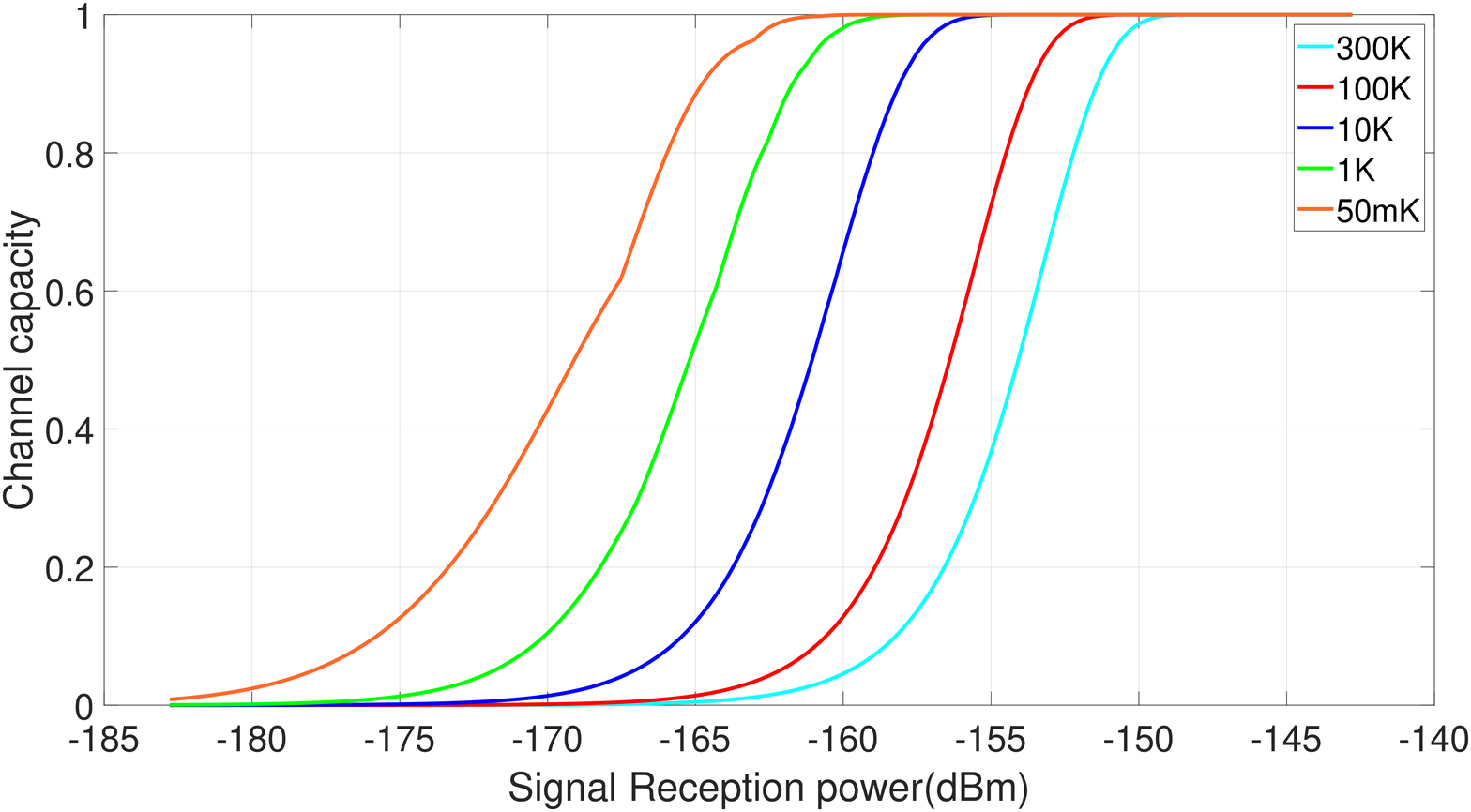}
  \caption{The achievable rate with respect to the signal power for different antenna temperatures.}
  \label{Likereal_x1}
\end{figure*}

Assume that the antenna is placed at temperature $T_A$, such that the background radiation energy is given by $kBT_A$ per slot; and the mean number of background photons per slot is $kBT_A/h\nu$. Assume that the signal reception power is $P_r$, such that the mean number of photons per unit time is $P_r/h\nu$. The achievable rates are shown in Figure \ref{Likereal_x1} for different temperatures $T_A$.


It can be seen that the achievable rate can reach above $0.95$ for signal power $-152$dBm at room temperature $300$K, and below $-160$dBm for temperature $50$mK. Considering the LTE signal sensitivity of $-100$dBm with data rate $2.2$Mbps, the power normalized to $1$kHz bandwidth is between $-130$dBm and $-135$dBm also at room temperature\cite{equipmentradio}. It is seen that, the sensitivity gain of our proposed structure is over $20$dB. Larger performance gain can be predicted via obtaining the achievable rate from two Poisson distributions without hard decision between them.

\section{MICROWAVE PHOTON DETECTION BASED ON MICROWAVE HBT EXPERIMENT}\label{sec.HBT}

We elaborate the output signal model of mircowave HBT experiment\cite{hbt}, show the equivalent communication system statistics, and numerically investigate the achievable rate.
\subsection{Microwave Hanbury Brown-Twiss Experiment\cite{hbt}}
\begin{figure}[htbp]
  \setlength{\abovecaptionskip}{-0.2cm} 
  \setlength{\belowcaptionskip}{-2cm}
  \centering
  \includegraphics[width=\columnwidth]{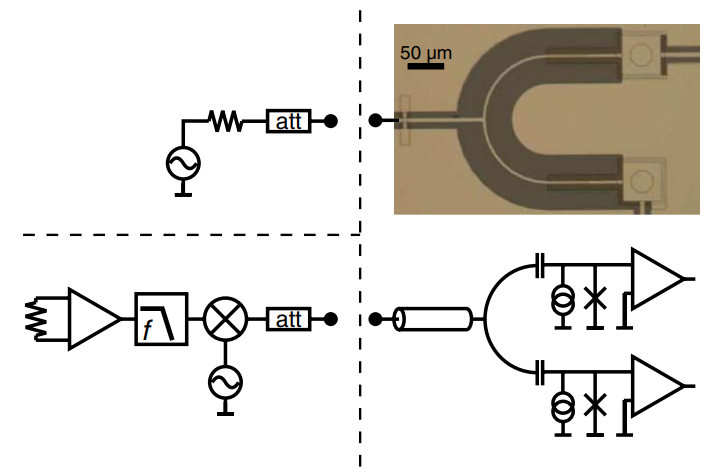}
  \caption{The HBT experimental devices,  where coherent microwave photon stream (upper left) and thermal noise microwave photon stream (bottom left) are divided into two-way counts (right side).\cite{hbt}}
  \label{hbt_exp}
\end{figure}

In the microwave HBT experiment, the microwave photons emitted by the source are divided into two paths, which are detected by two microwave photon counters with relative delay.
Both coherent microwave and thermal microwave are generated. The coherent microwave is generated by a microwave generator whose output is coupled to the detector through a transmission line with strong attenuation; and the thermal microwave is generated by amplifying, filtering, and upconverting the Johnson noise of the room temperature resistor. Double junction circuit is used to detect the statistics of the microwave photon source. Each junction has its own bias and readout. The schematic diagram of the experimental device is shown in Figure \ref{hbt_exp}.

\begin{figure}[htbp]
  \setlength{\abovecaptionskip}{-0.2cm} 
  \setlength{\belowcaptionskip}{-2cm}
  \centering
  \includegraphics[width=\columnwidth]{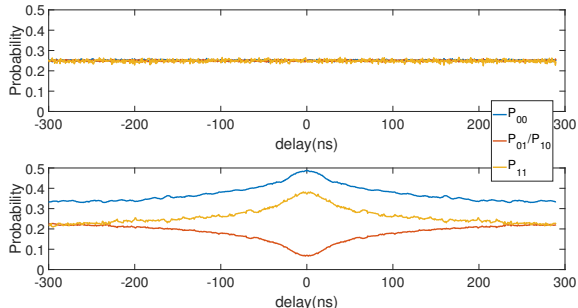}
  \caption{The probability with respect to different delays for coherent microwave (top) and thermal microwave (bottom).\cite{hbt}}
  \label{HBT_P}
\end{figure}

In the experiments, the output of two branches consists of two states, supercurrent state and voltage state, denoted as ``0'' and ``1'', respectively. Let $\tau$ denote the relative delay between the two measurements in the two branches. Let $P_{00}(\tau)$, $P_{01}(\tau)$, $P_{10}(\tau)$ and $P_{11}(\tau)$ denote the probability that the two branches are in states ``00'', ``01'', ``10'' and ``11'', respectively, for delay $\tau$ between the two branches. The values of $P_{00}(\tau)$, $P_{01}(\tau)$, $P_{10}(\tau)$ and $P_{11}(\tau)$ with respect to $\tau$ for coherent microwave and thermal microwave are shown in Figure~$\ref{HBT_P}$. It is seen that for coherent microwave, the probabilities are constant for different delay; while for thermal microwave the probabilities vary with relative delay $\tau$.

\subsection{Equivalent SIMO Channel Modeling}
Refer to the system model shown in Figure~$1$. Also assume OOK modulation and strong signal attenuation, such that the received signal cannot form continuous waveform. Within a symbol duration, the mean number of photons for symbol on is $\lambda_1$, and the mean number of background radiation is $\lambda_0$.

Assume that the signal paths are symmetrical, i.e, $P_{01}(\tau)$ = $P_{10}(\tau)$. For coherent sources, we have $P_{11}^{c}(\tau)$ = $p_{c}^{2}$,  $P_{10}^{c}(\tau)$ = $P_{01}^{c}(\tau)$ = $p_{c}(1-p_{c})$, and $P_{00}^{c}(\tau)$ = $(1-p_{c})^{2}$ for certain probability $p_c$; and for thermal noise, we have $P_{11}^{T}(\tau)+P_{10}^{T}(\tau)$ = $P_{T}$, $P_{10}^{T}(\tau)$ = $P_{01}^{T}(\tau)$.

Let $N_{00}^{c}, N_{10}^{c},$ and $N_{01}^{c}$ denote the number of pulse combinations corresponding to states $00, 10$, and $01$, respectively; and the counterparts for thermal noise are given by $N_{00}^{T}, N_{10}^{T},$ and $N_{01}^{T}$, respectively.

The number of photons actually received for states $00$, $10$ and $01$, denoted as $N_{00}$, $N_{10}$ and $N_{01}$, respectively, are given by
\begin{equation}
\begin{aligned} N_{00} &=N_{00}^{c}+N_{00}^{T}, \\ N_{10} &=N_{10}^{c}+N_{10}^{T}, \\ N_{01} &=N_{01}^{c}+N_{01}^{T}. \end{aligned}
\end{equation}

\begin{figure}[htbp]
  \setlength{\abovecaptionskip}{-0.2cm} 
  \setlength{\belowcaptionskip}{-2cm}
  \centering
  \includegraphics[width=\columnwidth]{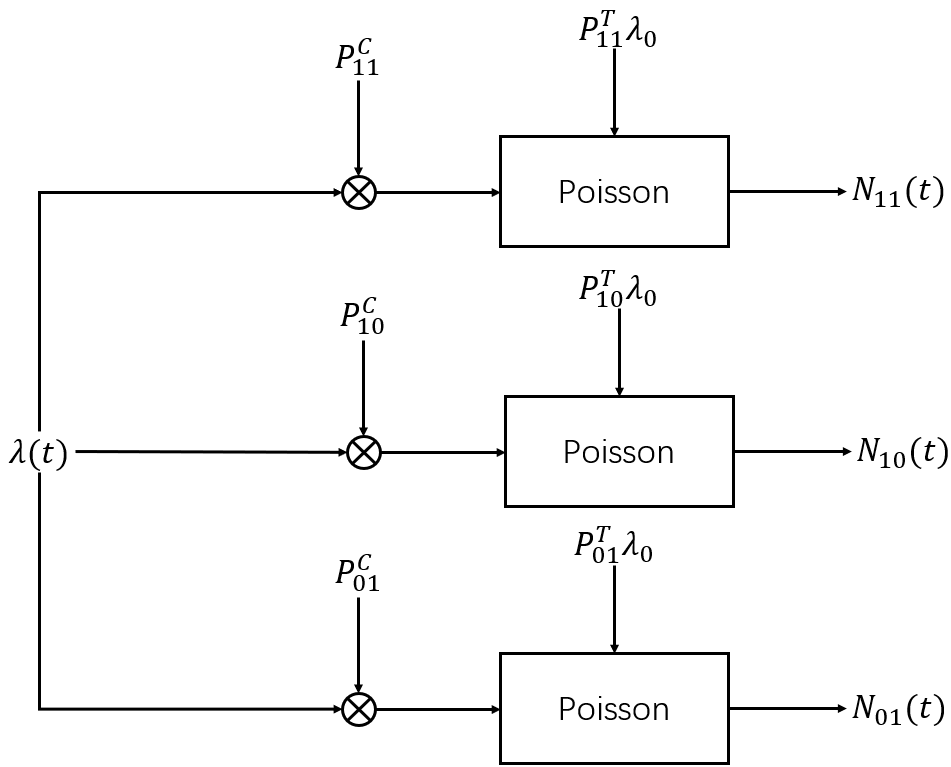}
  \caption{The quantumized microwave communication channel based on HBT experiment.}
  \label{poisson_hbt1}
\end{figure}

The channel model is shown in Figure \ref {poisson_hbt1}. Now we give the following probability distribution at the receiving side, for symbols one and zero

\begin{equation}
\begin{aligned}
\Pi_{1}(k_{00}, k_{01}, k_{10}) = p\left(N_{00}=k_{00}, N_{01}=k_{01}, N_{10}=k_{10} | 1\right) \\
=\prod_{i\in\{00,01,10\}} e^{-(P_{i}^{C}\lambda_1+P_{i}^{T}\lambda_0) } \frac{\left(P_{i}^{C}\lambda_1+P_{i}^{T}\lambda_0\right)^{k_{i}}}{k_{i} !},
\end{aligned}
\end{equation}
\begin{equation}
\begin{aligned}
\Pi_{0}(k_{00}, k_{01}, k_{10}) = p\left(N_{00}=k_{00}, N_{01}=k_{01}, N_{10}=k_{10} | 0\right)\\
=\prod_{i\in\{00,01,10\}} e^{-P_{i}^{T} \lambda_0} \frac{\left(P_{i}^{T} \lambda_0\right)^{k_{i}}}{k_{i} !}.
\end{aligned}
\end{equation}

Let $p$ denote the prior distribution of symbol on. The achievable rate via OOK modulation is given by
\begin{equation}
\begin{aligned}
C=&\max _{0 \leq p \leq 1} -\sum_{k_{00}, k_{01}, k_{10}} \Pi_{p}(k_{00}, k_{01}, k_{10})\log \Pi_{p}(k_{00}, k_{01}, k_{10})\\
&-p \Pi_{1}(k_{00}, k_{01}, k_{10}) \log \Pi_{1}(k_{00}, k_{01}, k_{10})\\
&-(1-p) \Pi_{0}(k_{00}, k_{01}, k_{10})\log \Pi_{0}(k_{00}, k_{01}, k_{10}),
\end{aligned}
\end{equation}
where $\Pi_{p}(k_{00}, k_{01}, k_{10})=p \Pi_{1}(k_{00}, k_{01}, k_{10})+(1-p) \Pi_{0}(k_{00}, k_{01}, k_{10})$. The second derivative of mutual information is
\begin{equation}
\frac{\partial^{2} I}{\partial^{2} p}=-\sum \frac{\left(\Pi_{1}(k_{00}, k_{01}, k_{10})-\Pi_{0}(k_{00}, k_{01}, k_{10})\right)^{2}}{\Pi_{p}(k_{00}, k_{01}, k_{10})} \leq 0,
\end{equation}
The optimal prior probability $p$ can be solved by convex optimization.

\subsection{Channel Simulation}
Assume that the symbol duration is $1$ms, and that the coherent photon frequency $\mathrm{\nu}=3.8 \mathrm{GHz}$\cite{hbt}. We take typical coherent photon absorption probability $\left[P_{00}^{C}(\tau), P_{10}^{C}(\tau), P_{01}^{C}(\tau)\right]=[0.25\ 0 .25\ 0 .25]$, and thermal noise photon absorption probability $\left[P_{00}^{T}(\tau), P_{10}^{T}(\tau), P_{01}^{T}(\tau)\right]=[0.36\ 0 .07\ 0 .07]$. Other simulation configurations are the same as those in Section \ref{sec.SAB_sum}.

\begin{figure*}[htbp]
  \setlength{\abovecaptionskip}{-0.2cm} 
  \setlength{\belowcaptionskip}{-2cm}
  \centering
  \includegraphics[width=2\columnwidth]{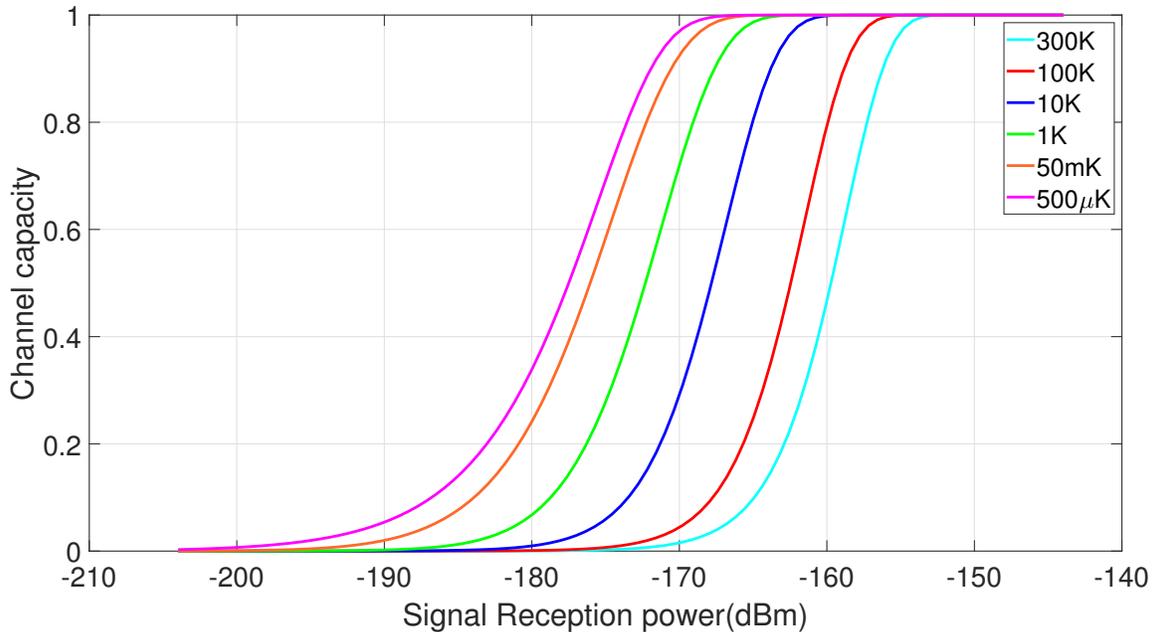}
  \caption{The achievable rate with respect to the signal power for different antenna temperatures.}
  \label{Likereal_hbt1}
\end{figure*}

Assume that the antenna is placed at temperature $T_A$, such that the background radiation energy is given by $kBT_A$ per slot; and the mean number of background photons per slot is $kBT_A/h\nu$. The achievable rates are shown in Figure \ref{Likereal_hbt1} for different temperatures $T_A$.

It is seen that the achievable rate can reach above $0.95$ for signal power $-156$dBm at room temper ature $300$K, and below $-170$dBm for temperature $50$mK. From Section \ref{sec.SAB_sum}, the normalized power of LTE signal to $1$kHz bandwidth is at room temperature is between $-130$dBm to $-135$dBm. Then, the performance gain of our proposed approach is over $15$dB.
\section{CONCLUSION}\label{sec.CONCLUSION}
We have envisaged a wireless communication system architecture based on microwave photon-level detection with superconducting
devices. We have reviewed existing works on microwave photon sensing using superconducting structure, which can improve the receiver-side detection sensitivitycompared with the state-of-the-art semi-conducting devices. We have also adopted two types of microwave photon detection methods based on microwave absorption barrier and HBT experiments, characterized the corresponding communication channel models and predicted the achievable communication rate. It is seen that the microwave photon-level signal detection can improve the receiver sensitivity compared with the state-of-the-art communication system with waveform signal reception, with gain over $10$dB. According to the performance prediction, our system can achieve significant performance improvement in low temperature environments. At present, commercial compact refrigerators can readily reduce the ambient temperature from 300K to below 10K within a few hours. So it is possible to deploy microwave photon-level detectors working in a low-temperature environment to increase the communication sensitivity.

In future work, we will focus on the coding based on the communication system architecture in Section \ref{sec.sys}, and further communication performance improvement brought by more advanced superconducting devices. Moreover, conducting real experiments to evaluate the real communication system performance comprises of another major endeavor in the future.

\small{\baselineskip = 10pt
	\bibliographystyle{IEEEtran}
    \bibliography{ustc}

\begin{thebibliography}{10}
\providecommand{\url}[1]{#1}
\csname url@samestyle\endcsname
\providecommand{\newblock}{\relax}
\providecommand{\bibinfo}[2]{#2}
\providecommand{\BIBentrySTDinterwordspacing}{\spaceskip=0pt\relax}
\providecommand{\BIBentryALTinterwordstretchfactor}{4}
\providecommand{\BIBentryALTinterwordspacing}{\spaceskip=\fontdimen2\font plus
\BIBentryALTinterwordstretchfactor\fontdimen3\font minus
  \fontdimen4\font\relax}
\providecommand{\BIBforeignlanguage}[2]{{%
\expandafter\ifx\csname l@#1\endcsname\relax
\typeout{** WARNING: IEEEtran.bst: No hyphenation pattern has been}%
\typeout{** loaded for the language `#1'. Using the pattern for}%
\typeout{** the default language instead.}%
\else
\language=\csname l@#1\endcsname
\fi
#2}}
\providecommand{\BIBdecl}{\relax}
\BIBdecl

\bibitem{sis}
J.~R. Tucker and M.~J. Feldman, ``Quantum detection at millimeter
  wavelengths,'' \emph{Rev. Mod. Phys.}, vol.~57, pp. 1055--1113, 1985.

\bibitem{hamidun}
M.~H. Cohen, L.~M. Falicov, and J.~C. Phillips, ``Superconductive tunneling,''
  \emph{Phys. Rev. Lett.}, vol.~8, pp. 316--318, Apr 1962.

\bibitem{SIS1}
C.~A. Mears, Q.~Hu, P.~L. Richards, A.~H. Worsham, D.~E. Prober, and A.~V.
  Räisänen, ``Quantum‐limited heterodyne detection of millimeter waves
  using superconducting tantalum tunnel junctions,'' \emph{Applied Physics
  Letters}, vol.~57, no.~23, pp. 2487--2489, 1990.

\bibitem{SIS2}
P.~L. {Richards} and Q.~{Hu}, ``Superconducting components for infrared and
  millimeter-wave receivers,'' \emph{Proceedings of the IEEE}, vol.~77, no.~8,
  pp. 1233--1246, 1989.

\bibitem{SIS3}
M.~J. {Wengler}, N.~{Dubash}, G.~{Pance}, and R.~E. {Miller}, ``Josephson
  effect gain and noise in sis mixers,'' \emph{IEEE Transactions on Microwave
  Theory and Techniques}, vol.~40, no.~5, pp. 820--826, 1992.

\bibitem{SIS4}
M.~J. {Wengler}, ``Submillimeter-wave detection with superconducting tunnel
  diodes,'' \emph{Proceedings of the IEEE}, vol.~80, no.~11, pp. 1810--1826,
  1992.

\bibitem{guangji1}
L.~Midolo, A.~Schliesser, and A.~Fiore, ``Nano-opto-electro-mechanical
  systems,'' \emph{Nature Nanotechnology}, vol.~13, no.~1, pp. 11--18, 2018.

\bibitem{guangji2}
A.~P. Higginbotham, P.~S. Burns, M.~D. Urmey, R.~W. Peterson, N.~S. Kampel,
  B.~M. Brubaker, G.~Smith, K.~W. Lehnert, and C.~A. Regal, ``Harnessing
  electro-optic correlations in an efficient mechanical converter,''
  \emph{Nature Physics}, vol.~14, no.~10, pp. 1038--1042, 2018.

\bibitem{guangji3}
S.~Barzanjeh, M.~Abdi, G.~J. Milburn, P.~Tombesi, and D.~Vitali, ``Reversible
  optical-to-microwave quantum interface,'' \emph{Phys. Rev. Lett.}, vol. 109,
  p. 130503, 2012.

\bibitem{guangji4}
K.~Zhang, F.~Bariani, Y.~Dong, W.~Zhang, and P.~Meystre, ``Proposal for an
  optomechanical microwave sensor at the subphoton level,'' \emph{Phys. Rev.
  Lett.}, vol. 114, p. 113601, 2015.

\bibitem{guangji5}
O.~c.~v. \ifmmode~\check{C}\else \v{C}\fi{}ernot\'{\i}k, S.~Mahmoodian, and
  K.~Hammerer, ``Spatially adiabatic frequency conversion in
  optoelectromechanical arrays,'' \emph{Phys. Rev. Lett.}, vol. 121, p. 110506,
  2018.

\bibitem{guangji6}
R.~W. Andrews, R.~W. Peterson, T.~P. Purdy, K.~Cicak, R.~W. Simmonds, C.~A.
  Regal, and K.~W. Lehnert, ``Bidirectional and efficient conversion between
  microwave and optical light,'' \emph{Nature Physics}, vol.~10, pp. 321 EP --,
  Mar 2014.

\bibitem{NV1}
T.~Joas, A.~M. Waeber, G.~Braunbeck, and F.~Reinhard, ``Quantum sensing of weak
  radio-frequency signals by pulsed mollow absorption spectroscopy,''
  \emph{Nature Communications}, vol.~8, no.~1, p. 964, 2017.

\bibitem{NV2}
B.~{Yang}, Y.~{Dong}, Z.~{Hu}, G.~{Liu}, Y.~{Wang}, and G.~{Du}, ``Noninvasive
  imaging method of microwave near field based on solid-state quantum
  sensing,'' \emph{IEEE Transactions on Microwave Theory and Techniques},
  vol.~66, no.~5, pp. 2276--2283, 2018.

\bibitem{counter}
R.~J. {Schoelkopf}, S.~H. {Moseley}, C.~M. {Stahle}, P.~{Wahlgren}, and
  P.~{Delsing}, ``A concept for a submillimeter-wave single-photon counter,''
  \emph{IEEE Transactions on Applied Superconductivity}, vol.~9, no.~2, pp.
  2935--2939, 1999.

\bibitem{dianlv}
G.~de~Lange, B.~van Heck, A.~Bruno, D.~J. van Woerkom, A.~Geresdi, S.~R.
  Plissard, E.~P. A.~M. Bakkers, A.~R. Akhmerov, and L.~DiCarlo, ``Realization
  of microwave quantum circuits using hybrid superconducting-semiconducting
  nanowire josephson elements,'' \emph{Phys. Rev. Lett.}, vol. 115, p. 127002,
  2015.

\bibitem{exp1}
N.~Bergeal, F.~Schackert, L.~Frunzio, and M.~H. Devoret, ``Two-mode correlation
  of microwave quantum noise generated by parametric down-conversion,''
  \emph{Phys. Rev. Lett.}, vol. 108, p. 123902, 2012.

\bibitem{exp2}
A.~V. Semenov, I.~A. Devyatov, P.~J. de~Visser, and T.~M. Klapwijk, ``Coherent
  excited states in superconductors due to a microwave field,'' \emph{Phys.
  Rev. Lett.}, vol. 117, p. 047002, 2016.

\bibitem{hbt}
Y.-F. Chen, D.~Hover, S.~Sendelbach, L.~Maurer, S.~T. Merkel, E.~J. Pritchett,
  F.~K. Wilhelm, and R.~McDermott, ``Microwave photon counter based on
  josephson junctions,'' \emph{Phys. Rev. Lett.}, vol. 107, p. 217401, 2011.

\bibitem{airtoforce}
S.~Barzanjeh, M.~Abdi, G.~J. Milburn, P.~Tombesi, and D.~Vitali, ``Reversible
  optical-to-microwave quantum interface,'' \emph{Phys. Rev. Lett.}, vol. 109,
  p. 130503, 2012.

\bibitem{airtoelc}
S.~Barzanjeh, S.~Guha, C.~Weedbrook, D.~Vitali, J.~H. Shapiro, and
  S.~Pirandola, ``Microwave quantum illumination,'' \emph{Phys. Rev. Lett.},
  vol. 114, p. 080503, 2015.

\bibitem{info1sec3}
A.~D. {Wyner}, ``Capacity and error exponent for the direct detection photon
  channel-part i-ii,'' \emph{IEEE Transactions on Information Theory}, vol.~34,
  no.~6, p. 1449–1471, 1988.

\bibitem{info2sec3}
K.~{Chakraborty}, S.~{Dey}, and M.~{Franceschetti}, ``Outage capacity of mimo
  poisson fading channels,'' \emph{IEEE Transactions on Information Theory},
  vol.~54, no.~11, pp. 4887--4907, 2008.

\bibitem{info3sec3}
A.~{Lapidoth} and S.~M. {Moser}, ``On the capacity of the discrete-time poisson
  channel,'' \emph{IEEE Transactions on Information Theory}, vol.~55, no.~1,
  pp. 303--322, 2009.

\bibitem{ISI}
C.~{Gong} and Z.~{Xu}, ``Channel estimation and signal detection for optical
  wireless scattering communication with inter-symbol interference,''
  \emph{IEEE Transactions on Wireless Communications}, vol.~14, no.~10, pp.
  5326--5337, 2015.

\bibitem{ISI4}
------, ``Non-line of sight optical wireless relaying with the photon counting
  receiver: A count-and-forward protocol,'' \emph{IEEE Transactions on Wireless
  Communications}, vol.~14, no.~1, pp. 376--388, 2015.

\bibitem{XX3sec3a}
C.~{Gong}, K.~{Wang}, Z.~{Xu}, and X.~{Wang}, ``On full-duplex relaying for
  optical wireless scattering communication with on-off keying modulation,''
  \emph{IEEE Transactions on Wireless Communications}, vol.~17, no.~4, pp.
  2525--2538, 2018.

\bibitem{gujic1}
K.~{Wang}, C.~{Gong}, D.~{Zou}, and Z.~{Xu}, ``Turbulence channel modeling and
  non-parametric estimation for optical wireless scattering communication,''
  \emph{Journal of Lightwave Technology}, vol.~35, no.~13, pp. 2746--2756,
  2017.

\bibitem{jiesh1}
D.~{Zou}, C.~{Gong}, K.~{Wang}, and Z.~{Xu}, ``Characterization on practical
  photon counting receiver in optical scattering communication,'' \emph{IEEE
  Transactions on Communications}, vol.~67, no.~3, pp. 2203--2217, 2019.

\bibitem{jiesh3}
G.~Wang, K.~Wang, C.~Gong, D.~Zou, Z.~Jiang, and Z.~Xu, ``A 1mbps real-time
  nlos uv scattering communication system with receiver diversity over 1km,''
  \emph{IEEE Photonics Journal}, 02 2018.

\bibitem{xsq}
G.~Romero, J.~J. Garc\'{\i}a-Ripoll, and E.~Solano, ``Microwave photon detector
  in circuit qed,'' \emph{Phys. Rev. Lett.}, vol. 102, p. 173602, 2009.

\bibitem{oneroad}
C.~Eichler, D.~Bozyigit, C.~Lang, L.~Steffen, J.~Fink, and A.~Wallraff,
  ``Experimental state tomography of itinerant single microwave photons,''
  \emph{Phys. Rev. Lett.}, vol. 106, p. 220503, 2011.

\bibitem{Pf1}
E.~C.~G. Sudarshan, ``Equivalence of semiclassical and quantum mechanical
  descriptions of statistical light beams,'' \emph{Phys. Rev. Lett.}, vol.~10,
  pp. 277--279, 1963.

\bibitem{Pf2}
R.~J. Glauber, ``Coherent and incoherent states of the radiation field,''
  \emph{Phys. Rev.}, vol. 131, pp. 2766--2788, 1963.

\bibitem{Qfunction}
K.~HUSIMI, ``Some formal properties of the density matrix,'' \emph{Proceedings
  of the Physico-Mathematical Society of Japan. 3rd Series}, vol.~22, no.~4,
  pp. 264--314, 1940.

\bibitem{tworoad}
E.~P. Menzel, F.~Deppe, M.~Mariantoni, M.~A. Araque~Caballero, A.~Baust,
  T.~Niemczyk, E.~Hoffmann, A.~Marx, E.~Solano, and R.~Gross, ``Dual-path state
  reconstruction scheme for propagating quantum microwaves and detector noise
  tomography,'' \emph{Phys. Rev. Lett.}, vol. 105, p. 100401, 2010.

\bibitem{equipmentradio}
U.~Equipment, ``radio transmission and reception (release 12), 3gpp ts 36.101,
  v12. 13.0, 2016.''

\end{thebibliography}

\end{document}